\documentclass[iop]{emulateapj}

\def\lya{Ly$\alpha$ }

\def\kms{{\rm km\, s^{-1}}}

\slugcomment{Draft, 20140722}

\begin{document}

\title{Anisotropic Lyman-alpha Emission}
\author{ Zheng Zheng
and
Joshua Wallace }
\affil{
 Department of Physics and Astronomy, University of Utah,
 115 South 1400 East, Salt Lake City, UT 84112
}
\begin{abstract}
As a result of resonant scatterings off hydrogen atoms, 
\lya emission from star-forming galaxies provides a probe of the
(hardly isotropic) neutral gas environment around them.
We study the effect of the environmental anisotropy on the
observed \lya emission by performing radiative transfer
calculations for models of neutral hydrogen clouds with prescriptions of 
spatial and kinematic anisotropies. The environmental anisotropy leads to 
corresponding anisotropy in the \lya flux and spectral properties and induces 
correlations among them. The \lya flux (or observed luminosity) depends on 
the viewing angle and shows an approximate correlation with the initial 
\lya optical depth in the viewing direction relative to those in 
all other directions. The distribution of \lya flux from a set of randomly 
oriented clouds is skewed to high values, providing a natural contribution 
to the \lya equivalent width (EW) distribution seen in observation. 
A narrower EW distribution is found at a larger peak offset of the 
\lya line, similar to the trend suggested in observation. The peak offset 
appears to correlate 
with the line shape (full width at half maximum and asymmetry), pointing to 
a possibility of using \lya line features alone to determine the systemic 
redshifts of galaxies. The study suggests that anisotropies in the 
spatial and kinematic distributions of neutral hydrogen can be an important 
ingredient in shaping the observed properties of \lya emission from 
star-forming galaxies. We discuss the implications of using \lya emission to 
probe the circumgalactic and intergalactic environments of galaxies.
\end{abstract}


\section{Introduction}

\lya emission from reprocessed ionizing photons in star-forming galaxies is
a prominent feature that has been used to detect high redshift galaxies 
\citep[e.g.,][]{Partridge67,Rhoads03,Gawiser07,Ouchi08,Hill08,Guaita10,
Ciardullo12}. 
Such \lya emitting galaxies (or \lya 
emitters; LAEs) have become an important laboratory to study galaxy 
formation, large-scale structure, and cosmic reionization. The resonant 
scattering of \lya photons with neutral hydrogen atoms in the circumgalactic
and intergalactic media (CGM and IGM) also potentially opens a window to probe
the spatial and kinematic environments of CGM and IGM from \lya emission.
In this paper,
we perform a theoretical study of the effect of anisotropy in the environment 
on \lya emission properties and study the implications of such an effect in
using \lya emission to probe galaxy environment.

Anisotropic gas distributions are common in galaxies and their surrounding 
environments. The anisotropy can show up both spatially and kinematically.
The gas distribution around star-forming regions (likely clumps on
a galaxy disk) is already anisotropic. The galactic winds from star formation
are ubiquitous and typically show a bipolar outflow pattern 
\citep[e.g.,][]{Bland88,Shopbell98,Veilleux02,Shapley03,Weiner09,Rubin13}. 
The cold gas being
supplied for star formation could be accreted from streams and filaments 
\citep[e.g.,][]{Keres05,Dekel09}.
Even on the IGM scale, the density field and the velocity field still show 
appreciable fluctuations \citep[e.g.,][]{Zheng10}. That is, spatial and
 kinematic anisotropies can exist in galactic environments on all scales.
 The resonant scatterings of
\lya photons would enable them to explore such anisotropies and encode 
information in the \lya emission properties.

The anisotropic gas density and velocity distributions are naturally produced 
in hydrodynamic simulations of galaxy formation. Monte Carlo \lya radiative 
transfer calculation has been performed for individual simulated galaxies. 
For example,
\citet{Laursen09} notice the anisotropic escape of \lya emission in nine
simulated galaxies.  
\citet{Barnes11} show that emerging \lya spectra from three simulated galaxies
vary strongly with the viewing angle.
The escape fraction of \lya emission in the simulated galaxies
in \citet{Yajima12} exhibits strong dependence on galaxy morphology and
orientation, and for a disk galaxy the escaping \lya photons are confined 
to a direction perpendicular to the disk.
\citet{Verhamme12} also find that \lya properties strongly depend on the 
disk orientation in one simulated galaxy with an outflowing velocity field.

As a statistical study, \citet{Zheng10} perform \lya radiative transfer 
calculation for about $2\times 10^5$ sources in a radiative hydrodynamic 
cosmological reionization simulation. Given the resolution of the simulation,
the density and velocity anisotropies come from gas in CGM and IGM.
The \lya emission properties (luminosity, surface brightness, and spectra) 
are found to depend on the viewing
angle, as a result of the density and velocity distributions (environment).
Because of the environment dependent radiative transfer, the \lya emission 
properties show correlations among themselves, and new effects in the spatial
clustering of LAEs are induced (\citealt{Zheng11a}; also see  
\citealt{Wyithe11,Behrens13}). 

On the analytic side, solutions are usually found 
for simple configurations of uniform media, e.g., static plane-parallel 
uniform slabs 
\citep[][]{Harrington73,Neufeld90}, static uniform spheres 
\citep[][]{Dijkstra06}, and uniform media experiencing Hubble expansion 
(in the diffusion regime; 
\citealt{Loeb99}). Numerical solutions or simulations for analytical setups 
are also usually focused on uniform slabs \citep[e.g.,][]{Auer68,Avery68,
Adams72,Ahn00,Ahn01,Ahn02}, or uniform spheres or isotropic systems 
\citep[e.g.,][]{Loeb99,Ahn02b,Zheng02,Dijkstra06,Verhamme06,Roy09,Roy10}.

Given the potential importance of the anisotropy in the observed 
properties of LAEs, it is useful to investigate 
\lya emission with systems of analytic setups incorporating prescriptions
of anisotropy. Such an investigation can guide our analyses of \lya emission
from simulated galaxies and help our understanding of the role of
anisotropy in shaping the \lya emission from LAEs. In this paper, we 
perform such a study.

The structure of the paper is as follows.
In \S~\ref{sec:model}, we describe how we build models of neutral
hydrogen clouds with spatial or kinematic anisotropy for \lya radiative
transfer calculation. Then we present the results of anisotropies in the
escaping \lya emission from these models in \S~\ref{sec:results}.
Finally, in \S~\ref{sec:summary} we summarize our investigation and 
discuss the implications on studying \lya emission from star-forming galaxies.

\section{Models of Anisotropic Neutral Hydrogen Clouds}
\label{sec:model}

To investigate the effect of anisotropic density and velocity distribution 
on the \lya emission, we construct three simple models of spherical neutral
hydrogen gas clouds. The first two models intend to 
investigate the effects of anisotropy induced by density and velocity 
separately. The third one is motivated by galactic wind and mimics an outflow
confined in a cone. 
For all the three models, the temperature of the neutral hydrogen atoms in 
each cloud is fixed at $2\times 10^4$~K. The \lya emitting source is assumed 
to be a point source located at the cloud center.

We emphasize that while our models may capture different aspects of the 
gas distribution around galaxies, by no means they are realistic. The purpose 
of the study is to investigate the effects from the anisotropy in the two main
quantities that affects \lya radiative transfer, i.e., density and velocity. 
Instead of more sophisticated models with various couplings between density
and velocity, we intentionally separate them and build simple models to see 
the effect from each component. 

The first model we consider is a ``density gradient'' case. In this model,  
there is no bulk motion of the gas in the cloud and the anisotropic optical 
depth distribution is purely a result of the anisotropy in the density 
distribution. Specifically, we introduce a density gradient along the $z$ 
direction on top of an otherwise uniform cloud. The neutral hydrogen number
density $n(z)$ of the cloud is parameterized as
\begin{equation}
\label{eqn:dengrad}
n(z)=\bar{n}\left(1-2A\frac{z}{R}\right),
\end{equation}
where $\bar{n}$ is the mean number density in the cloud, $R$ is the cloud 
radius, and $A$ is a parameter denoting the magnitude of the density 
gradient. The column density from the cloud center follows a dipole 
distribution $N_{\rm HI}=\bar{n}R(1-A\cos\theta)$, where $\theta$ is the angle 
with respect to the $+z$ direction (i.e., the polar angle).

The optical depth of \lya photons depends not only on the density distribution
but also on the velocity distribution. Photons of the same frequency
appear to have different frequency shifts in the restframe of atoms for atoms 
with different velocities in the lab frame (i.e., the Doppler effect), which 
leads to different 
probability of interacting with the atoms (i.e., different scattering 
cross-section). Therefore, for the second model, we consider a ``velocity
gradient'' case, where the difference in optical depth along different
directions is purely a result of the anisotropy in the velocity distribution.
With a uniform density cloud that can undergo uniform Hubble-like 
expansion/contraction, we introduce a velocity gradient along the $z$ 
direction,
\begin{equation}
\label{eqn:vgrad}
{\mathbf v}({\mathbf r})=\frac{r}{R} V \hat{\mathbf r} + \frac{z}{R} \Delta V \hat{z}_+.
\end{equation}
The first term on the right-hand side is the uniform Hubble-like motion, and
the second term represents the modification from the velocity gradient, with 
$\hat{\mathbf r}$ and $\hat{z}_+$ being the unit vectors along ${\mathbf r}$ 
and the $+z$ directions, respectively. The quantity $V$ is the Hubble velocity 
at the edge of the 
cloud and $\Delta V$ denotes the magnitude of the velocity gradient. A positive
(negative) value of $\Delta V$ means that an additional outflow (inflow) field
is added on both the $+z$ and $-z$ sides of the cloud.

Finally, motivated by galactic wind, we consider a case named ``bipolar wind''.
For a cloud of uniform density, we set up a Hubble-like velocity field within a
limited solid angle,
\begin{equation}
{\mathbf v}({\mathbf r}) = \left\{
 \begin{array}{ll}
   \frac{r}{R} V \hat{\mathbf r} & \quad \text{if $|z|/r>\mu_0$},\\
                                   & \\
   0                               & \quad \text{otherwise}.\\
 \end{array}\right.
\end{equation}
That is, the ``wind'' has a half open angle of $\theta_0=\cos^{-1} \mu_0$ 
around the $z$ axis. The model shares some similarity with the ``velocity 
gradient'' model. The differences are that here the velocity gradient is 
imposed along the radial direction (not $z$ direction) and the velocity field 
is confined to a cone. The geometry is similar to the model in 
\citet{Noterdaeme12}.

Our models can be regarded as simplistic models of star-forming galaxies 
emitting \lya photons with the cloud representing the CGM and IGM environments.
With the above models, we consider three cases of mean \lya optical depth. 
This can be put in terms of a characteristic column density, $N_{\rm HI}
=\bar{n}R$. The three cases have $N_{\rm HI}=10^{18}$, $10^{19}$, and $10^{20} 
{\rm cm}^{-2}$, ranging from \lya limit systems to damped \lya systems.

For each setup, we perform the radiative transfer calculation of \lya photons
using a Monte Carlo code \citep{Zheng02}. \lya photons are initially launched
isotropically from the cloud center, with frequency following a Gaussian 
profile 
centered at the rest-frame \lya line-center frequency with the width determined
by the temperature of the gas. For each run, we use $5\times 10^5$--$10^6$ 
photons to obtain good statistics on fluxes and spectra.

\lya photons are collected once they reach the surface of the cloud. 
We define the angle $\Theta$ as the polar angle of the escaped photon 
with respect to $+z$ axis as seen by a distant observer. Since this angle 
is the most observationally relevant one (and since all three models are 
axial symmetric about the $z$ axis), we focus our study on the \lya 
emission properties as a function of $\Theta$. For describing the cloud
configuration (e.g., density and velocity distribution), we use $\theta$ 
for the polar angle in the cloud's local frame, i.e., the angle between 
a given radial direction and the $+z$ axis.

\section{Results of the Radiative Transfer Calculation}
\label{sec:results}

Given the three types of models, different magnitudes of the density/velocity 
gradient, and different column densities, there are a large number of radiative 
transfer runs we perform. To obtain a basic picture of the anisotropic \lya 
emission, we start with the ``density gradient'' case with $N_{\rm HI}=10^{19}
{\rm cm^{-2}}$ and introduce our analyses in \lya flux and spectra. We then
present the ``velocity gradient'' case with $N_{\rm HI}=10^{19}
{\rm cm^{-2}}$ and the ``bipolar wind'' case. Finally we discuss the general 
features in the anisotropic \lya emission based on all the runs.

\subsection{``Density Gradient'' Case with $N_{\rm HI}=10^{19} {\rm cm^{-2}}$ }

\begin{figure*}
\plotone{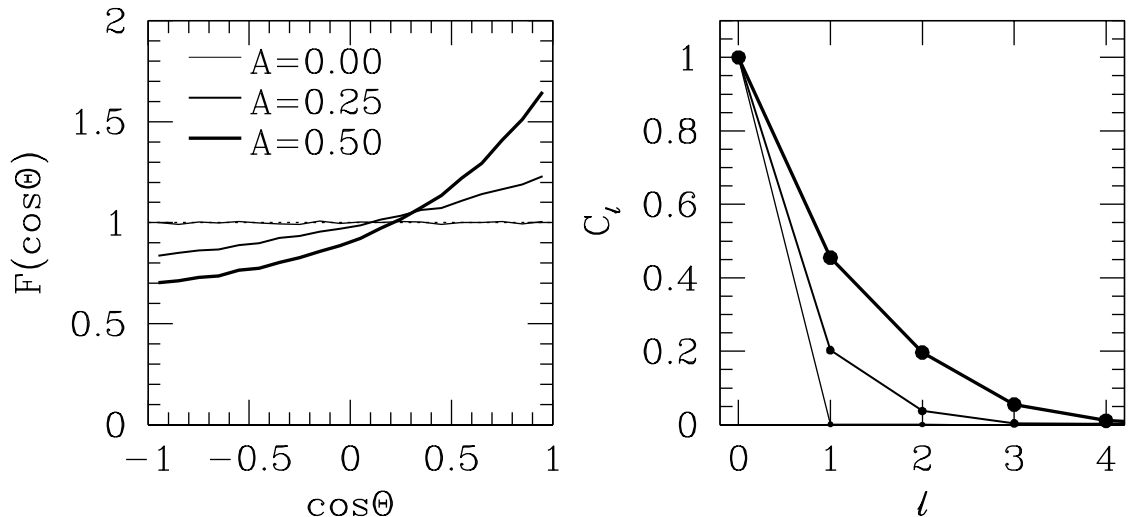}
\caption[]{
\label{fig:surfaceflux}
Distribution of \lya flux observed by distant observers for a static spherical 
cloud with
anisotropic density distribution (the ``density gradient'' case with column 
density of $10^{19}{\rm cm}^{-2}$). 
{\it Left panel}: \lya flux as a function of
the polar angle $\Theta$ of the escaping photons. Parameter
$A$ denotes the magnitude of the density gradient imposed along the $z$ 
direction, and the flux is normalized with respect to the isotropic flux
of a uniform cloud (i.e., the $A=0$ case). {\it Right panel}: the multipole
expansion coefficients of the anisotropic distribution of flux. See the text
for more details.
}
\end{figure*}

We first study the anisotropic \lya flux seen by distant observers.
The flux at a polar angle $\Theta$ is 
proportional to the number of photons $\Delta N$ in a narrow angular bin 
$\Delta\Theta$ divided by the corresponding area 
$2\pi D^2\sin\Theta\Delta\Theta$ for observers at distance $D$. 
In what follows, we will normalize this 
flux to the isotropic flux 
$N/(4\pi D^2)$. The normalized flux is then 
\begin{equation}
\label{eqn:F}
F(\mu)=\frac{2\Delta N}{N\Delta\mu},
\end{equation}
where $\mu=\cos\Theta$.
It can also be put as the fractional photon count 
$\Delta N/N$ divided by the fractional area $\Delta\mu/2$ around the given 
polar angle. It satisfies the normalization condition 
$\int_{-1}^1 F(\mu) d\mu/2 =1$.

The left panel of Figure~\ref{fig:surfaceflux} shows the angular dependence 
of the flux and the dependence on the magnitude of the density gradient (with
$A=0$ being the uniform sphere case). 
Resonant scatterings of \lya photons off neutral hydrogen atoms enable them 
to probe the optical depths along all directions, and they tend to 
preferentially make their way out along the path of least resistance. From
our setup of the ``density gradient'' case (Equation~\ref{eqn:dengrad}), 
the density decreases toward the $+z$ direction. For \lya photons at the
cloud center, the scattering optical depth is lowest (highest) along
the $\Theta=0$ ($\Theta=\pi$) direction. It can be clearly seen from the
figure that \lya photons favor escape along the $\Theta=0$ direction and
dislike escape along the $\Theta=\pi$ direction. The ratio of fluxes at 
$\Theta=0$ and at $\Theta=\pi$ increases as we increase the density gradient, 
reaching a factor of about 2.5 for $A=0.5$.

To quantify the anisotropy in the full range of the polar angle $\Theta$, we 
decompose the flux $F(\mu)$ into its multipole components,
\begin{equation}
F(\mu)=\sum_{l=0}^{\infty} C_l P_l(\mu),
\end{equation}
where $P_l$ is the $l$-th order Legendre polynomial.
The coefficient $C_l$ is solved from the orthogonality of the Legendre 
polynomials,
\begin{equation}
C_l=\frac{2l+1}{2}\int_{-1}^1 F(\mu)P_l(\mu) d\mu 
   =\frac{2l+1}{N} \sum_{i=1}^{N} P_l(\mu_i),
\end{equation}
where $\mu_i$ corresponds to the direction of the $i$-th photon 
($i=$1, 2, ..., $N$). The rightmost expression reduces the integral to a
simple sum over all photons. It is derived by making use of 
Equation~\ref{eqn:F} in the limit of infinitesimally small $\Delta \mu$ bin,
and in such a limit, $\Delta N$ is either 1 or 0.  The monopole coefficient 
$C_0$ of $F(\mu)$ is unity by definition.

The density gradient introduces a dipole component in the density 
distribution (and thus the initial optical depth distribution seen from the 
center). A corresponding dipole component in the \lya flux shows up
(right panel of Figure~\ref{fig:surfaceflux}). However, since 
\lya photons encounter a large number of scatterings and continually 
change their
travel directions and frequencies before emerging at the 
surface, their final distribution does not follow the initial optical depth 
distribution. As a result, higher multipole components emerge, and
their relative contribution increases as the density gradient increases.

\begin{figure*}
\plotone{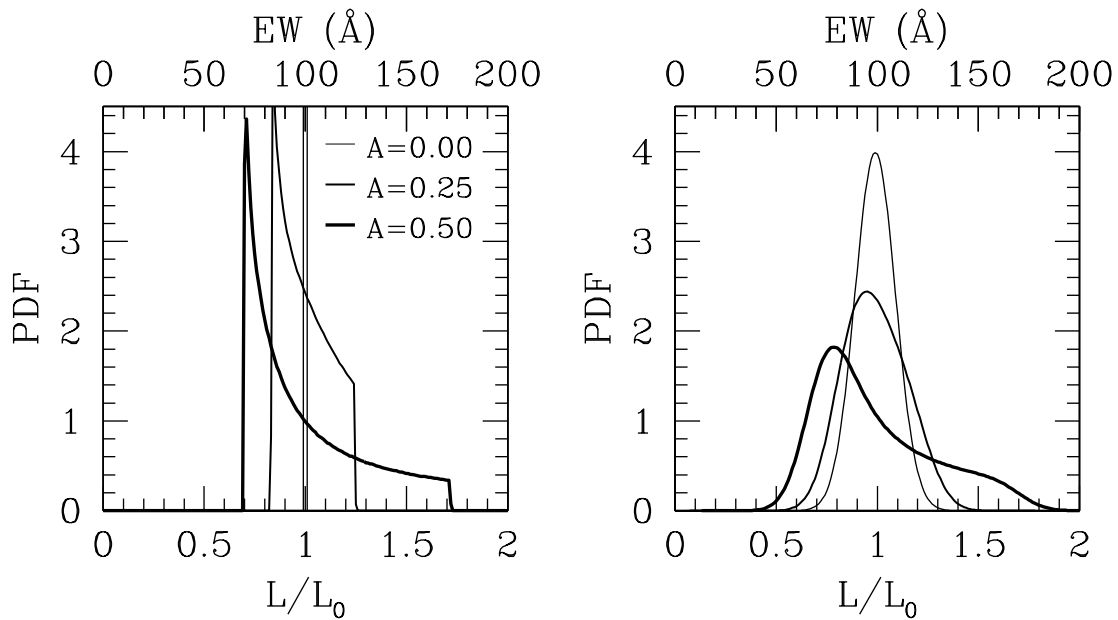}
\caption[]{
\label{fig:EW}
 Distribution of apparent (observed) \lya luminosity from observations along 
random directions of a static spherical cloud with
anisotropic density distribution 
(the ``density gradient'' case with column density 
$10^{19}{\rm cm}^{-2}$). The luminosity $L$ is in 
units of the intrinsic luminosity ($L_0$). Effectively it can be put in terms
of the \lya EW. In the top axis of each panel, the values of EW are marked
by assuming the intrinsic EW to be 100\AA\ (corresponding to a stellar 
population with Salpeter IMF and 1/20 solar metallicity with age above 100Myr;
see the text). The left panel shows the original distributions, while the 
right panel shows the ones smoothed with a Gaussian kernel with standard 
deviation of 10\AA\ to mimic the effect of measurement errors.
}
\end{figure*}

\begin{figure*}
\plotone{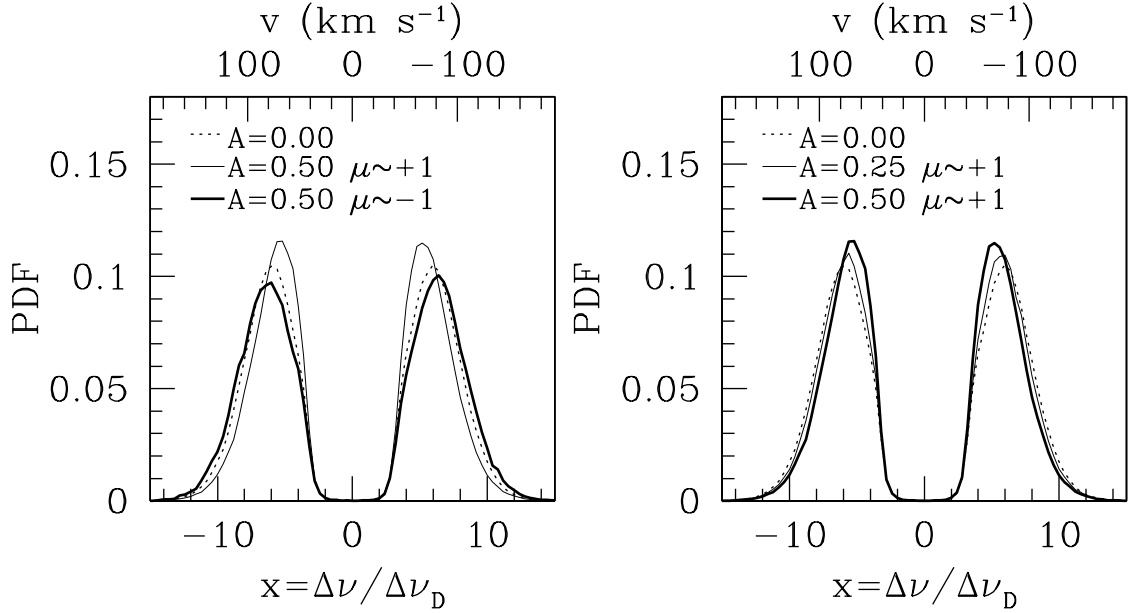}
\caption[]{
\label{fig:spec}
Normalized \lya spectra from a static spherical cloud with anisotropic density 
distribution (the ``density gradient'' case with column density 
$10^{19}{\rm cm}^{-2}$). {\it Left panel}: comparison of spectra observed
along the two pole directions ($\mu=\cos\Theta=\pm 1$) for the $A=0.50$ 
model. The directions with the lowest and highest column density are 
$\mu=+1$ and $\mu=-1$, respectively (see Equation~\ref{eqn:dengrad}). 
{\it Right panel}:
comparison of spectra observed along one pole direction for clouds with
different anisotropic parameter $A$. In both panels, the spectra from 
the uniform case ($A=0$) are shown for reference. The frequency offset in 
units of the Doppler parameter $\Delta\nu_D$ and in velocity units are
shown in the bottom and top axes, respectively.
}
\end{figure*}

\begin{figure*}
\plotone{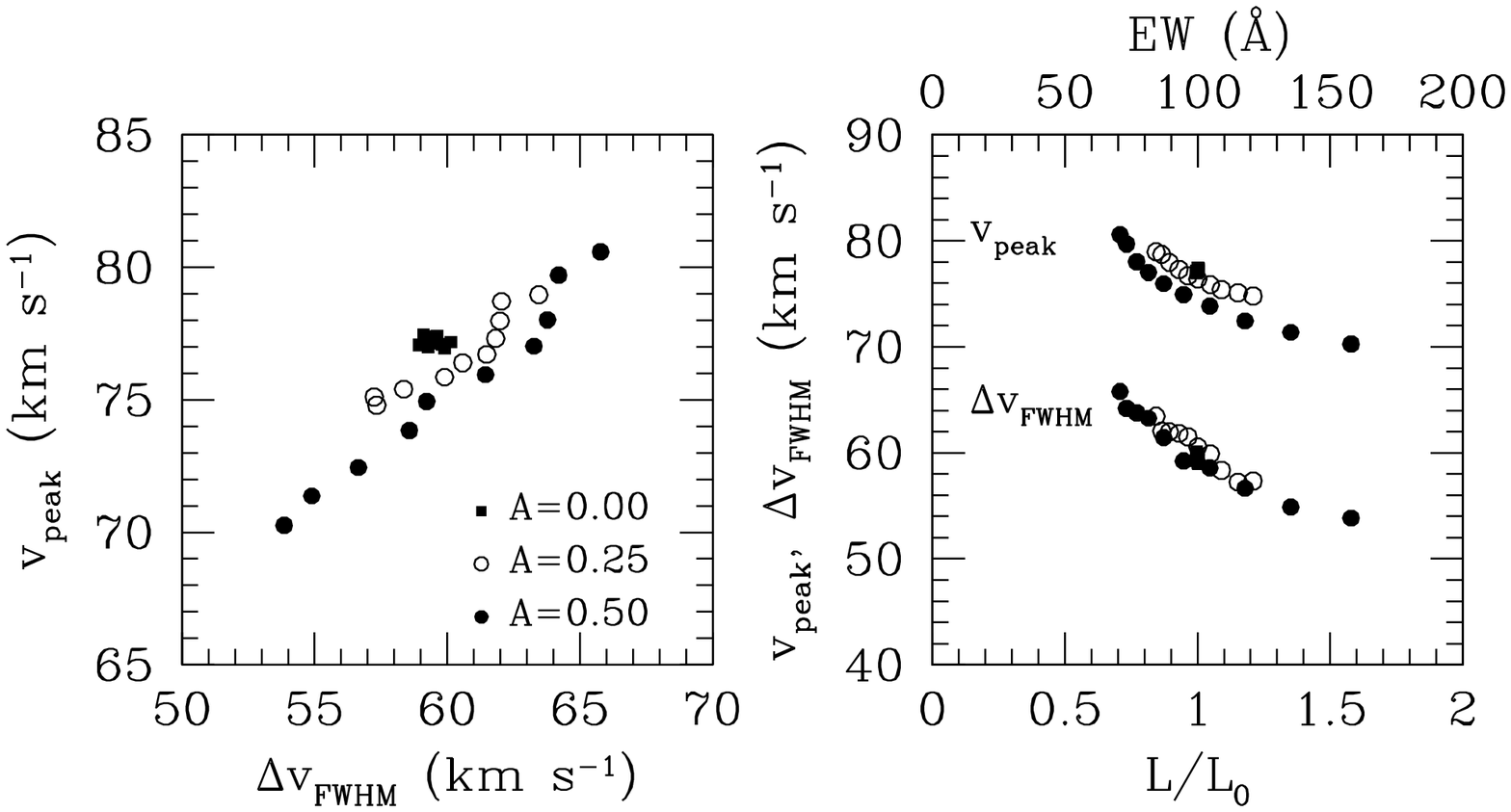}
\caption[]{
 \label{fig:Lspec}
 Peak offset $v_{\rm peak}$ and FWHM $\Delta v_{\rm FWHM}$ of \lya emission 
from a static spherical cloud with anisotropic density distribution (for the 
``density gradient'' case with column density $10^{19}{\rm cm}^{-2}$). 
 {\it Left panel}: the correlation between $v_{\rm peak}$ and 
 $\Delta v_{\rm FWHM}$. {\it Right panel}: their correlation with the
 apparent \lya luminosity or \lya EW. Only the red peak is analyzed here. 
}
\end{figure*}

Because of the anisotropy, the observed flux depends on the viewing angle
(or observed direction). Observers along different directions would infer
different luminosities of the source (assuming isotropic emission at the 
source). 
Equivalently, for a given observer, observations of similar clouds at
random orientations would give the distribution of the observed (apparent) 
luminosity. In Figure~\ref{fig:EW}, we show the distribution of the ratio
of apparent luminosity $L$ to intrinsic luminosity $L_0$ as a function of 
density gradient magnitude. The ratio $L/L_0$ is calculated in the same way 
as $F(\mu)$ in Equation~\ref{eqn:F}. The apparent luminosity is connected to
the equivalent width (EW) of \lya emission. If we neglect other contributions
to EW, like dust effect on the continuum and \lya emission 
\citep[e.g.,][]{Verhamme12} and observed \lya emission being only a fraction 
of the total \lya emission \citep[e.g.,][]{Zheng11b}, $L/L_0$ is proportional 
to the \lya EW. Therefore, the anisotropic emission 
provides a mechanism for the distribution of \lya EW. 

To be specific and for the convenience of the comparison, here we neglect 
other contributions and assume that the intrinsic \lya EW is 100\AA, 
corresponding to a stellar population with Salpeter initial stellar mass 
function (IMF) and 1/20 solar metallicity with age above 100Myr 
\citep[][]{Malhotra02}. The
top axis in each panel marks the resultant EW=$100{\rm \AA}(L/L_0)$. Note
that the $L/L_0$ (EW) distribution is computed based on the \lya flux 
distribution shown in the left panel of Figure~\ref{fig:surfaceflux}. To
reduce the effect of noise in the flux distribution, we use the multipole
expansion of the distribution up to to $l=4$ to derive the $L/L_0$ (EW)
distribution shown in Figure~\ref{fig:EW}.

In the left panel of Figure~\ref{fig:EW}, the original EW distributions from
different runs are shown. If the cloud is uniform ($A=0$), the EW does not
depend on viewing angle and we always have the intrinsic one (i.e., the 
distribution is just a Dirac $\delta$ function). As the cloud becomes 
anisotropic,
the viewing angle dependent \lya emission makes the EW distribution extended.
The distribution is skewed, with a higher amplitude at lower EW. The skewness
increases as the anisotropy becomes stronger (as a result of the larger density
gradient), and at $A=0.5$ the distribution shows a prominent tail towards high
EW values. Interestingly, such a shape of the EW distribution is similar to 
the observed ones for LAEs \citep[e.g.,][]{Ouchi08,Nilsson09,
Ciardullo12}. The 
model EW distribution has sharp edges (corresponding to the boundary of 
$\Theta=0$ and $\pi$), which are not seen in real observations. To make a 
better 
comparison, we need to take into account the uncertainties in measuring
the \lya EW in observation. In the right panel of Figure~\ref{fig:EW}, each
distribution is smoothed with a Gaussian kernel with standard deviation of
10\AA, roughly the size of the measurement errors in \lya EW from observation 
\citep[e.g.,][]{Ouchi08}. The skewness seen in the left panel remains after
smoothing, and for strong anisotropic cases, the EW distribution mimics those
seen in observation.

Besides the viewing angle dependent flux or apparent luminosity, the \lya 
spectra are also affected by the system anisotropy. In the left panel of
Figure~\ref{fig:spec}, the normalized spectra observed from observers 
located on the $+z$ ($\Theta=0$ or $\mu=1$) and $-z$ ($\Theta=\pi$ or 
$\mu=-1$) directions for the $A=0.50$ case are 
compared. The spectra for the isotropic case ($A=0$) is shown for reference.
The bottom axis shows the frequency offset in units the Doppler frequency
$\Delta\nu_D\equiv (v_p/c)\nu_0$, which corresponds to the frequency offset 
of a line-center photon (with frequency $\nu_0$) seen by an atom moving with 
the most probable thermal velocity $v_p=\sqrt{2kT/m_H}$. The top axis marks
the offset in velocity units.
To escape a static medium, \lya photons need to shift either to the blue
or red wing, where the optical depth is small. This results in a characteristic
double-peak profile \citep[e.g.,][]{Neufeld90,Zheng02}, which applies to
the spectra plotted here.

The two opposite directions in the left panel of Figure~\ref{fig:spec} 
correspond to the directions of minimum and maximum column density. 
In general, photons that escape from a direction with a lower column density
diffuse less in frequency space. As a consequence, the separation of the
two peaks is smaller and the width of each peak is narrower. In the right
panel, the spectra in the $\Theta=0$ direction are compared for the $A=0.25$
and $A=0.50$ cases. The $A=0.50$ case has the lower line-of-sight column 
density, which has narrower peaks with a closer separation. The spectra do 
not differ substantially from the uniform case (dotted curve). The reason is
that the column density modulation on top of the uniform case is only 
a factor of $1-A\cos\theta$. Even for the $A=0.5$ case, the modulation is only
a factor of three. Because of the resonant scattering, \lya photons can be
thought to probe the optical depths in all the directions before escaping along
the final direction. This also reduces the difference in the effective column 
densities experienced by photons escaping along different directions, leading 
to only small differences in the spectra.

The left panel of Figure~\ref{fig:Lspec} shows the viewing angle dependence
of two spectral features, the peak offset $v_{\rm peak}$ with respect to the 
restframe \lya line center and the line width characterized by the 
full width at half maximum (FWHM) $\Delta v_{\rm FWHM}$. In order to reduce
the effect of noise, a Gaussian fit around the peak in
each spectrum is made to determine the peak offset.
Both $v_{\rm peak}$ and $\Delta v_{\rm FWHM}$ 
are computed in ten bins of the viewing angle, with the bottom-left 
and upper-right points corresponding to $\cos\Theta\sim +1$ and $-1$, 
respectively. We only show the case from the peak in the red side of the 
spectra -- in reality, the blue peak likely becomes insignificant because of
the scattering in the IGM with Hubble flow \cite[e.g.,][]{Dijkstra07,Zheng10,Laursen11}. The $A=0$ case is shown to give a 
sense of the noise in our calculation of these quantities from the spectra.
There is a clear correlation between $v_{\rm peak}$ and $\Delta v_{\rm FWHM}$
-- as \lya photons diffuse more in the frequency space, the spectrum as a whole
shifts further away from the initial line center and at the same time it 
becomes more broadened. 

Since both the apparent \lya luminosity and spectral features depend on viewing
angle, there must exist correlations between them. In the right panel of 
Figure~\ref{fig:Lspec}, we see that both the peak offset and width are 
anti-correlated with the \lya luminosity
(or EW). This makes perfect sense -- along a direction of easier escape, \lya
emission has a higher flux (thus larger $L/L_0$ or EW), while \lya photons 
diffuse less in frequency space (thus smaller peak offset and width).

As a whole, the resonant scatterings experienced by \lya photons constantly 
change their propagation directions. This enables the photons to probe optical 
depths along
all directions, and they tend to escape from the directions with low 
optical depth. For the ``density gradient'' case, the anisotropic distribution 
in the column density of the cloud translates to anisotropic 
\lya emission, i.e.,
viewing angle dependent \lya emission properties, such as the apparent \lya 
luminosity, spectral peak offset, and peak width. The viewing angle dependent 
\lya flux leads to a spread in \lya EW, suggesting an interesting
mechanism to produce the observed EW distribution. The emission properties are 
correlated as a result of the viewing angle dependence. In directions of lower
resistance, \lya photons tend to have higher flux and less diffusion in 
frequency space. 

\subsection{``Velocity Gradient'' Case with $N_{\rm HI}=10^{19}{\rm cm}^{-2}$}

\begin{figure*}[h]
\plotone{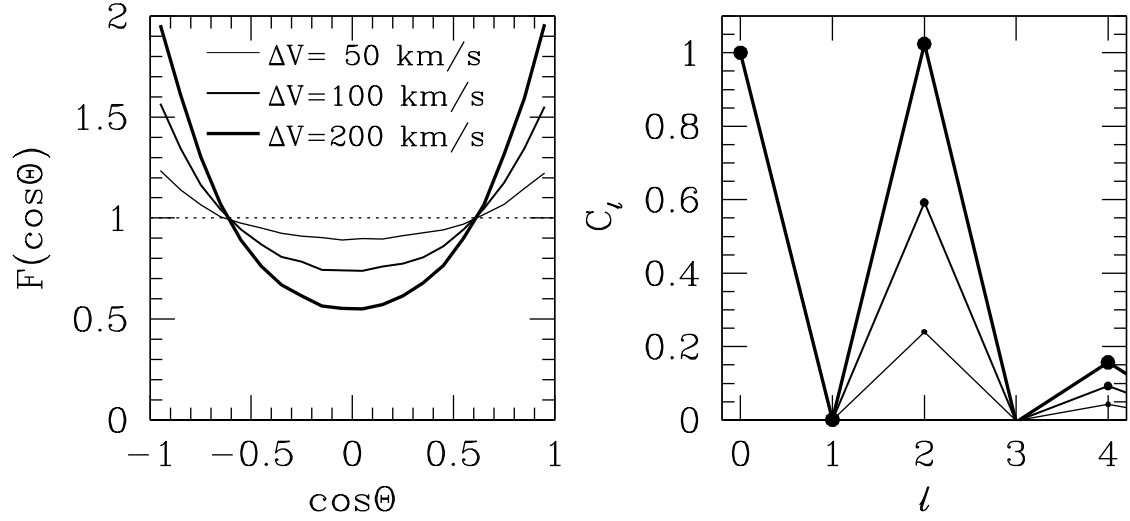}
\caption[]{
\label{fig:Vel_obsflux}
Similar to Figure~\ref{fig:surfaceflux}, but for a uniform density cloud with 
velocity anisotropy (the 
``velocity gradient'' case with column density $10^{19}{\rm cm}^{-2}$).
The polar angle $\Theta$ is the angle between direction of the distant 
observer and
the $z$ axis. The parameter $\Delta V$, the expansion velocity at the cloud 
poles, denotes the magnitude of the velocity anisotropy.
See text for more details.
}
\end{figure*}

\begin{figure*}[h]
\plotone{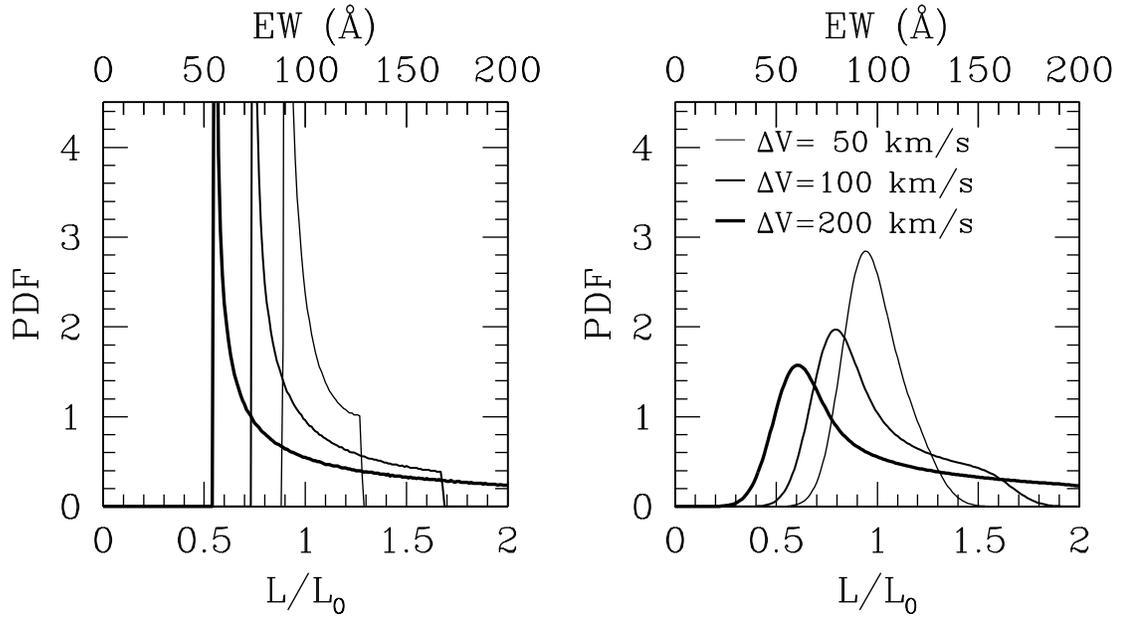}
\caption[]{
\label{fig:Vel_EW}
Similar to Figure~\ref{fig:EW}, but for a uniform density cloud with
velocity anisotropy (the ``velocity gradient'' case with column density
$10^{19}{\rm cm}^{-2}$).
The parameter $\Delta V$, the expansion velocity at the poles,
denotes the magnitude of the velocity anisotropy.
}
\end{figure*}

\begin{figure*}[h]
\plotone{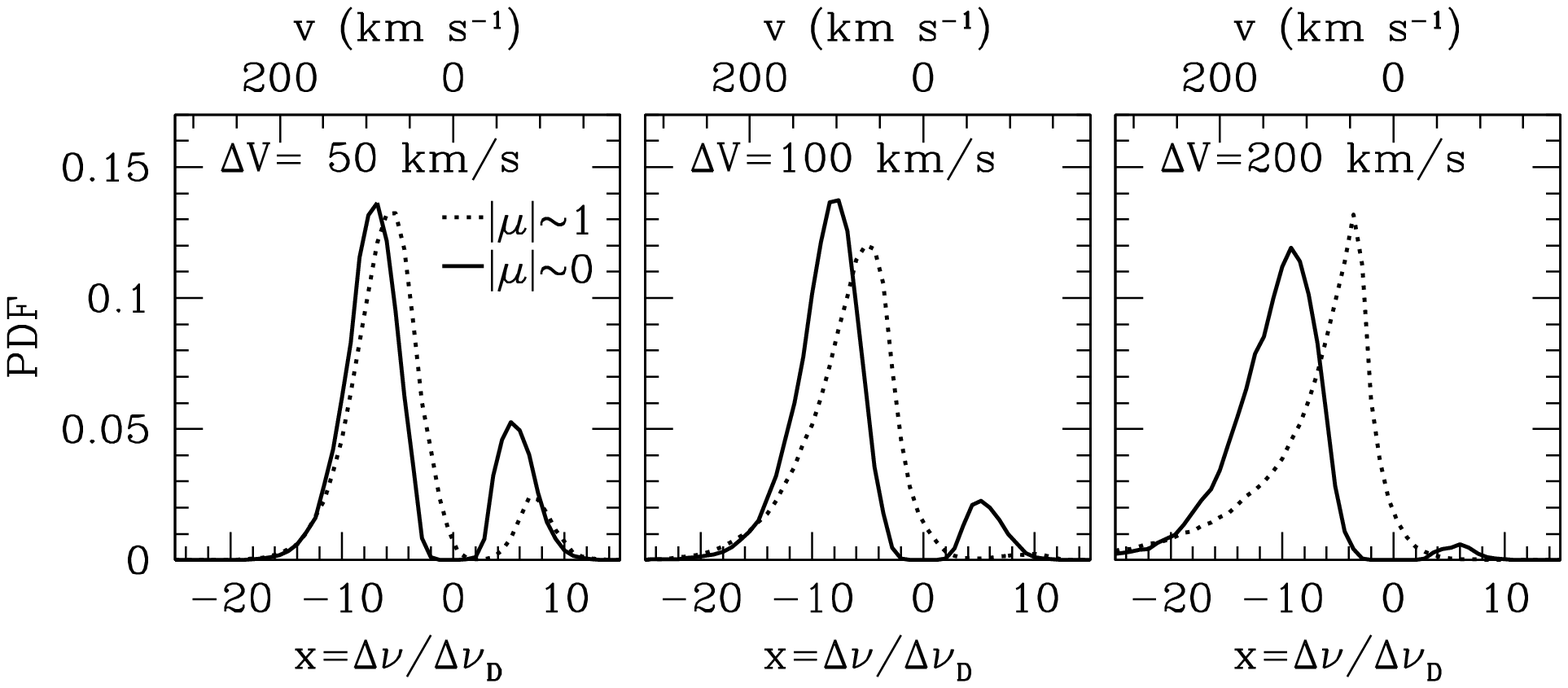}
\caption[]{
\label{fig:Vel_spec}
Similar to Figure~\ref{fig:spec}, but for a uniform density cloud with
velocity anisotropy (the ``velocity gradient'' case with column density 
$10^{19}{\rm cm}^{-2}$).
}
\end{figure*}

\begin{figure*}[h]
\plotone{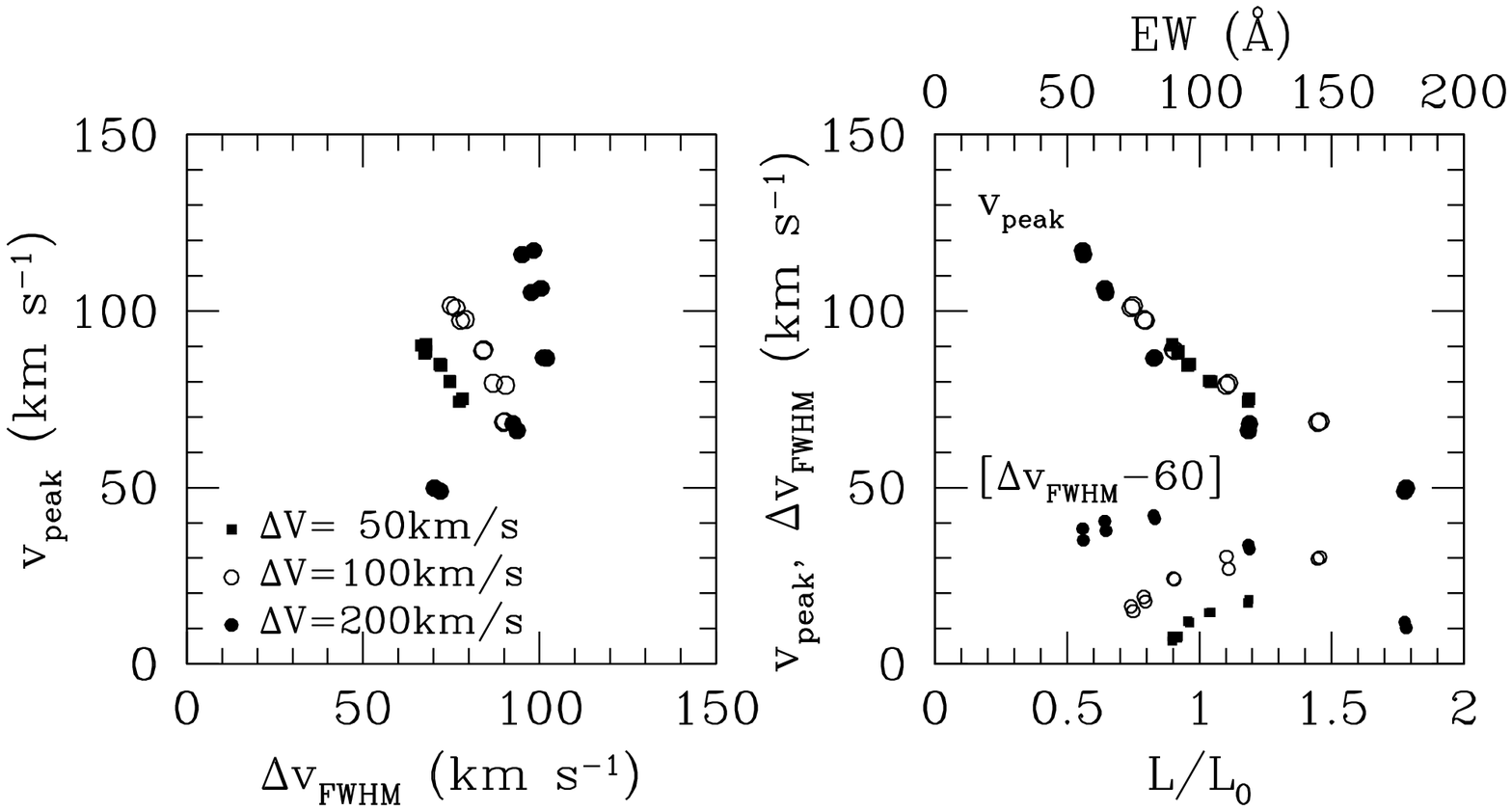}
\caption[]{
 \label{fig:Vel_Lspec}
Similar to Figure~\ref{fig:Lspec}, but for a uniform density cloud with
velocity anisotropy (the ``velocity gradient'' case with column density 
$10^{19}{\rm cm}^{-2}$).
}
\end{figure*}

Following the detailed discussions of the ``density gradient'' case,
we now turn to another anisotropic case by adding a velocity gradient to an
otherwise static uniform sphere ($V=0$ in Equation~\ref{eqn:vgrad}). The 
velocity gradient leads to anisotropy in the initial \lya optical depth, and 
we expect the \lya emission to depend
on viewing angle as well. As an example, we present the results for the 
$N_{\rm HI}=10^{19}{\rm cm}^{-2}$ case.

Figure~\ref{fig:Vel_obsflux} shows the viewing angle dependence of the 
flux measured by distant observers (left) and the multipole decomposition 
(right). The way we add the 
velocity gradient can be thought as Hubble-like expansion in the $\pm z$ 
directions and no expansion in the $\pm x$ and $\pm y$ directions. In the plot,
$\Delta V$ (taking the values of 50, 100, and 200 ${\rm km\, s^{-1}}$) is the 
radial velocity at the edge of the cloud along $\pm z$
directions. For such expansion velocities, the initial \lya photons launched
at the center are effectively at a single frequency (the restframe 
line-center frequency). In the restframe of the atoms in the expanding cloud, 
these initial photons appear to be redshifted off the line center, with
a frequency shift proportional to the radial velocity. With our setup,
the optical depth seen by the initial \lya photons appears low towards the
$\Theta=0$ and $\pi$ directions (poles) and high towards the $\Theta=\pi/2$
directions (equator). Therefore, unlike the ``density gradient'' case, here we
have a quadrupole-like distribution of the initial optical depth.

Given such a distribution of optical depth, it is expected that photons prefer
to escape towards the poles rather than the equator, as seen in the left 
panel of
Figure~\ref{fig:Vel_obsflux}. The ratio of the maximum to minimum fluxes 
increases from about two for $\Delta V=100\kms$ to about four for 
$\Delta V=200\kms$. A strong quadrupole component in the angular distribution 
of flux emerges (right panel of Figure~\ref{fig:Vel_obsflux}), which increases 
as $\Delta V$ increases. Unlike the ``density gradient'' case, there is no 
dipole component from the symmetry of the system. Nevertheless, the density and 
velocity gradient cases share the same effect on \lya emission -- anisotropy 
in the initial \lya optical depth leads to anisotropy in the \lya flux. 

Similar to the ``density gradient'' case, the anisotropic \lya emission
leads to a skewed distribution of apparent \lya luminosity ($L/L_0$ or EW) 
with a tail of high values (see Figure~\ref{fig:Vel_EW}, left for the original
distributions and right for the ones smoothed with a Gaussian kernel with 
a standard deviation of 10\AA). For $\Delta V$=100$\kms$ and 200$\kms$, the 
shape of the smoothed distribution looks similar to the observed ones
\citep[e.g.,][]{Ouchi08,Nilsson09,Ciardullo12}.

Figure~\ref{fig:Vel_spec} shows the spectra of photons 
escaping from the directions of the pole and equator as a function of the 
velocity gradient (parameterized by the velocity $\Delta V$ at the edge of 
the cloud). While the spectra in the ``velocity gradient'' case still have 
two peaks, they are no longer symmetric about the initial line center 
(Figure~\ref{fig:Vel_spec}). Blue \lya photons would be redshifted close to 
the line center in the restframe of hydrogen atoms in an expanding cloud and 
be strongly scattered. That is, \lya photons tend to 
shift redward for an easy escape from the cloud. Overall, the blue peak is
suppressed compared to the red peak. 
The generic features of the spectra can be understood
following the interpretations in e.g., \citet{Loeb99}, \citet{Zheng02},
\citet{Dijkstra06}, and \cite{Verhamme06}, by accounting for the anistropy.

Since both the density and velocity affect the frequency diffusion in the
``velocity gradient'' case, the relatively tight correlation between the peak 
position and peak FWHM seen in the ``density gradient'' case becomes weaker 
(left panel of Figure~\ref{fig:Vel_Lspec}). So does the correlation between 
apparent luminosity $L/L_0$ (or EW) and FWHM (lower points in the right panel).
However, the tight anti-correlation between $L/L_0$ (EW) and peak position 
$v_{\rm peak}$ persists (big symbols in the right panel). This anti-correlation
appears to be much stronger than that in the ``density gradient'' case 
(Figure~\ref{fig:Lspec}). 

To summarize, for ``velocity gradient'' induced optical depth anisotropy, \lya 
emission again displays the corresponding anisotropy in flux and spectral 
features. The apparent luminosity or EW distribution closely resembles the 
observed ones. As an extension, we also study systems with ``velocity 
gradient'' in one direction imposed on top of an isotropic Hubble-like 
expansion, and the relation
between EW and $v_{\rm peak}$ is similar to the ``velocity gradient'' cases
shown here. The results for the extended cases will be presented in 
\S~\ref{sec:general}.

\subsection{``Bipolar Wind'' Case}

\begin{figure*}
\plotone{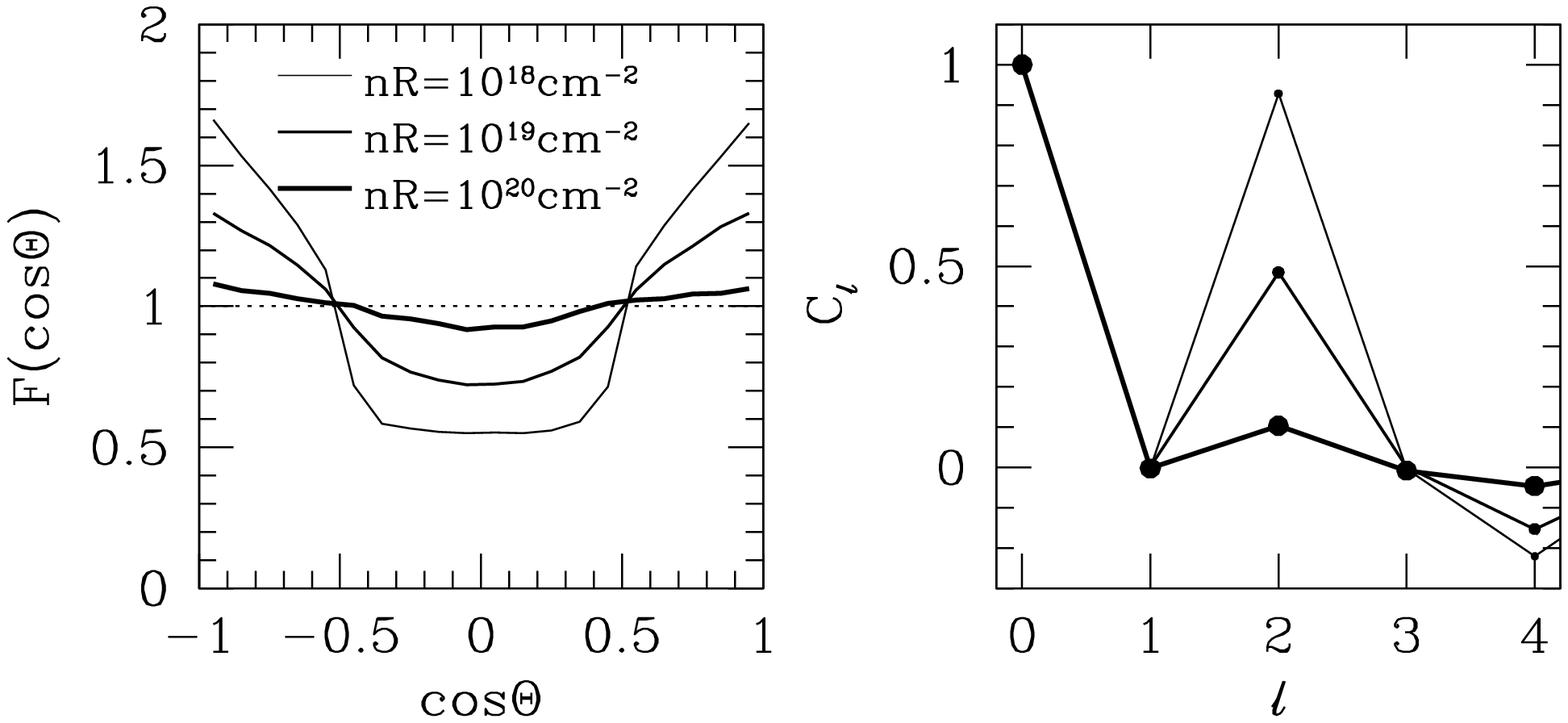}
\caption[]{
\label{fig:Bi_obsflux}
Similar to Figure~\ref{fig:Vel_obsflux},
but for a uniform density cloud with 
bipolar outflows (the ``bipolar wind'' case). 
}
\end{figure*}

Before discussing general properties of anisotropic \lya emission, we briefly
present the results on the flux anisotropy for the ``Bipolar Wind'' case. 
In this case, the uniform 
cloud has a Hubble-like expansion within a cone defined by 
$\theta<60\degr$ and $\theta>120\degr$ ($|\mu|>0.5$).
and remains static outside of the cone. That is, half of the cloud gas is 
expanding. The expansion velocity at the edge of the cloud is set to be 
200$\kms$. This specific setup is motivated by the bipolar outflow from 
galaxies. We perform runs for three column densities: $nR=10^{18}$, $10^{19}$,
and $10^{20}{\rm cm}^{-2}$. 

The \lya optical depth is determined by both the column density and velocity. 
For initial photons with frequency $\nu_i$, the radial optical depth is 
\begin{eqnarray}
\tau_{\rm ini}
     & = & \int_0^R 
n\sigma\left[\nu_i\left(1-\frac{\Delta V}{c}\frac{r}{R}\right)\right] dr
\nonumber\\
     & = & nR \int_0^1 \sigma\left[\nu_i(1-s\Delta V/c)\right] ds,
\end{eqnarray}
where $s=r/R$ and $\sigma$ is the \lya scattering cross-section. The equation 
applies to the regions both with and without outflow ($\Delta V\neq0$ 
and $\Delta V=0$). Since all the three runs have 
the same velocity fields, from the above equation we see that the 
above three runs have the same fractional anisotropy in the initial \lya 
optical depth and they differ in the overall optical depth scale (which is 
set by the column density).

The ``bipolar wind'' case shares some similarities with the
``velocity gradient'' case; the main differences are that its velocity 
field is radially oriented (rather than only in $\pm z$ direction) 
and is only within a limited region (rather than the whole cloud).
As a consequence, the results on the flux anisotropy, apparent luminosity 
distribution, and anisotropic spectral properties are also similar to the
``velocity gradient'' case. For brevity, we only present the flux 
anisotropy here and have other properties 
incorporated into the summary of the general results (\S~\ref{sec:general}).


Given the system setup, the anisotropy in 
the \lya emission (Figure~\ref{fig:Bi_obsflux}) 
mainly has a quadrupole component. As the column density
increases, \lya photons experience more scatterings and more changes in 
travel directions. The anisotropy in the initial \lya optical depth is
therefore becomes less important, and as a consequence the anisotropy in 
\lya flux becomes weaker.

\subsection{General Results}
\label{sec:general}

With the three cases discussed in the previous subsections, we turn to
present the general results and make an attempt to further connect to 
observations.

A few more model runs are included here. We extend the 
``velocity gradient'' case
by imposing a velocity gradient along the $z$ direction on top of a spherical
cloud with Hubble-like expansion (see Equation~\ref{eqn:vgrad}). The 
Hubble-like expansion has a velocity $V=100\kms$ at the edge of the cloud
and we impose different velocity gradients with the parameter $\Delta V$
ranging from -100 to 100$\kms$ with a step size 50$\kms$. We also perform 
runs with different column densities ($nR=10^{18}$, $10^{19}$, and $10^{20} 
{\rm cm^{-2}}$) for the ``density gradient'' case and the ``velocity 
gradient'' case and its extension. In total, the results of 32 runs from 
all the cases are presented here.

\begin{figure}
\plotone{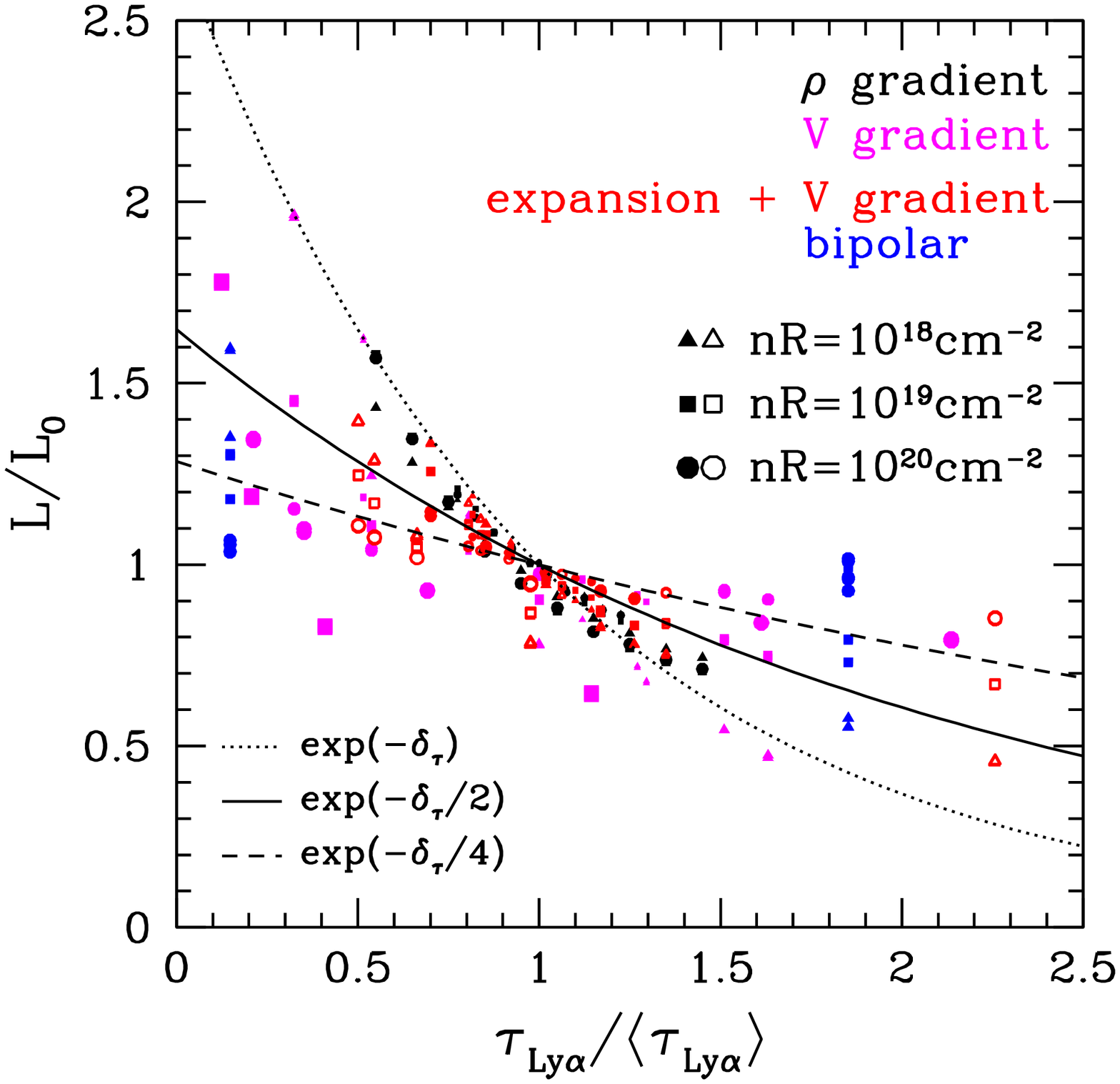}
\caption[]{
\label{fig:Ltau}
 The relation between the apparent \lya luminosity and the relative initial
 line-center optical depth. The value $\langle\tau_{\rm Ly\alpha}\rangle$
 is the line-center optical depth for initial \lya photons, averaged over
 all directions.
 The curves show $\exp(-\delta_\tau)$, 
 $\exp(-\delta_\tau/2)$, and $\exp(-\delta_\tau/4)$ to guide the eye,
 where $\delta_\tau$ is the optical depth excess defined as
 $\delta_\tau\equiv \tau_{\rm Ly\alpha}/\langle\tau_{\rm Ly\alpha}\rangle-1$.
 The anisotropy models include those caused by density anisotropy 
 (``density gradient'' case; black points),
by velocity anisotropy (``velocity gradient'' case; magenta points) and its
extension with an additional isotropic expansion component (``expansion +
velocity gradient''; red points), and by bipolar outflow (``bipolar wind'' case;
blue points). Systems with three different column densities are studied for
each case, $10^{18}$ (triangles), $10^{19}$ (squares), and
$10^{20} {\rm cm^{-2}}$ (circles). 
Open and filled symbols denote setups with negative and positive velocity 
gradient imposed, respectively.
}
\end{figure}

Overall, we find system anisotropy leads to anisotropies in \lya emission.
The observed \lya luminosity depends on the viewing angle. Since higher 
luminosities are likely observed along directions of easier escape, 
we expect the
luminosity to show a correlation with the optical depth of the same direction.
However, we do not expect the correlation to be tight. \lya photons probe 
the optical depth in all directions in a convoluted way, since their scattering 
with the neutral hydrogen atoms causes their positions, directions, and 
frequencies to continually change. Furthermore, for observed/apparent 
\lya luminosity, the absolute
value of optical depth is not a good indicator. As an example, consider the
uniform sphere case. For the same intrinsic luminosity, spheres of different
column density (optical depth) have the same observed luminosity, since all
\lya photons escape.\footnote{ 
This particular example shows that simply applying a $\exp(-\tau_\nu)$ 
correction to the initial \lya spectra is not the right way to do the \lya 
radiative transfer, where $\tau_\nu$ is the frequency dependent \lya optical
depth along the line-of-sight direction. This is demonstrated and emphasized
in \citet{Zheng10}.
}
A better quantity that connects to luminosity could be
the relative optical depth among all directions. 

\begin{figure}
\plotone{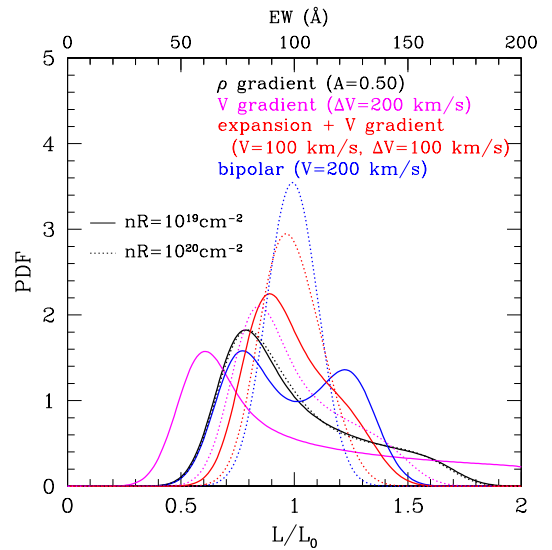}
\caption[]{
\label{fig:EWs}
Distribution of apparent (observed) \lya luminosity (or EW) from a few 
selected models. The curves have been smoothed with a Gaussian kernel with 
standard deviation of 10\AA\ to mimic the effect of measurement errors.
}
\end{figure}

In Figure~\ref{fig:Ltau},
we plot the observed luminosity $L$ in units of the intrinsic luminosity 
$L_0$ as a function of the relative optical depth 
$\tau_{\rm Ly\alpha}/\langle\tau_{\rm Ly\alpha}\rangle$ for all the models. 
Here $\tau_{\rm Ly\alpha}$ is the initial line-center optical depth for \lya 
photons and $\langle\tau_{\rm Ly\alpha}\rangle$ is the average over all 
directions. Or equivalently we can define the fractional excess 
$\delta_\tau=\tau_{\rm Ly\alpha}/\langle\tau_{\rm Ly\alpha}\rangle -1$
in the initial line-center optical depth.
The three curves are shown in the form of $\exp(-\delta_\tau)$, 
$\exp(-\delta_\tau/2)$, and $\exp(-\delta_\tau/4)$ to guide the eye.
Clearly there exists an overall correlation between apparent luminosity 
(or EW) and the relative line-center optical depth. The plot serves as a
nice summary of the anisotropic \lya emission from all the models considered
in this paper. It depicts the idea that \lya photons prefer to escape along 
the direction of least resistance. The resonant scatterings allow \lya photons
to probe optical depth along different directions, and the key quantity that 
determines the final anisotropic distribution of \lya emission is the relative 
optical depth, not the absolute one. 

As a result of the anisotropy, the observed \lya luminosity from a set of 
randomly oriented clouds will spread out around the intrinsic luminosity.
Figure~\ref{fig:EWs} summarizes the distributions of the apparent \lya
luminosity (EW) by showing the result from a few selected models. 
The apparent luminosity/EW distribution is largely determined by the 
anisotropy (in the initial optical depth) of the system. For a fixed 
anisotropy factor (e.g., density or velocity gradient), a cloud with higher 
column density has a lower degree of anisotropy in the initial optical 
depth distribution. This translates to smaller spread in the apparent 
luminosity/EW distribution (e.g., comparing the $10^{19} {\rm cm^{-2}}$ 
cases with those of $10^{20} {\rm cm^{-2}}$). The anisotropy leads to 
a skewed EW distribution with a tail toward higher values.
It becomes more skewed with higher degrees of flux anisotropy 
(i.e., lower column density). For the high column density ``bipolar wind'' 
case, a low amplitude bump is seen at higher EW values. For this model, 
the superposition of the EW distributions from systems with different column 
densities (more abundant for lower column density) can produce a distribution 
with an extended tail. The skewed EW distribution looks similar to the 
observed ones \citep[e.g.,][]{Ouchi08,Nilsson09,Ciardullo12}, which implies 
that anisotrpic \lya emission provides one mechanism to contribute to the 
\lya EW distribution. 

\begin{figure*}
\plotone{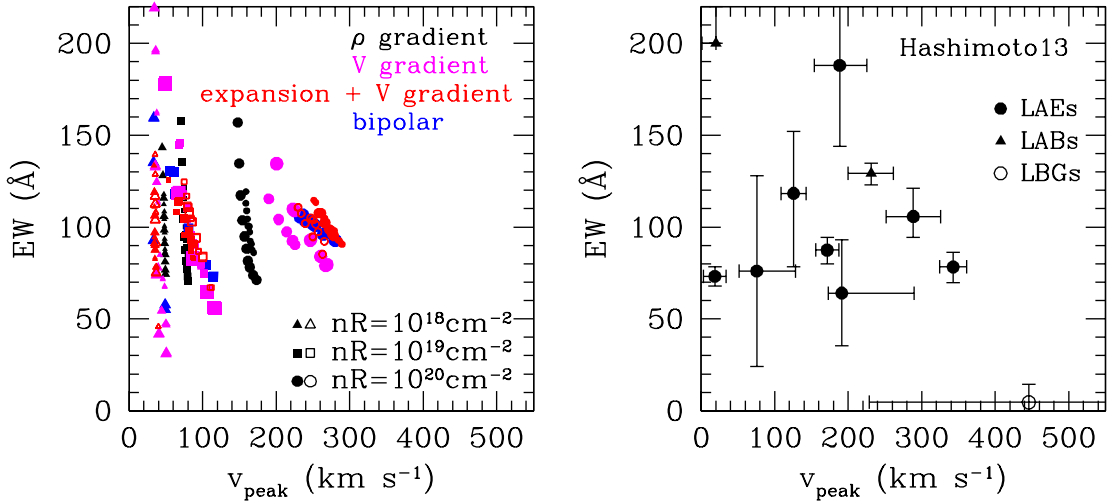}
\caption[]{
\label{fig:EW_Vpeak}
 The relation between the \lya EW and \lya line peak offset $v_{\rm peak}$.
{\it Left panel}: the relation from our models of anisotropic clouds, including
those caused by density anisotropy (``density gradient'' case; black points), 
by velocity anisotropy (``velocity gradient'' case; magenta points) and its 
extension with an additional isotropic expansion component (``expansion + 
velocity gradient''; red points), and by bipolar outflow (``bipolar wind'' case;
blue points). Systems with three different column densities are studied for
each case, $10^{18}$ (triangles), $10^{19}$ (squares), and 
$10^{20} {\rm cm^{-2}}$ (circles). At a given column density of each case,
the symbol size indicates the degree of anisotropy of the system (larger
for stronger anisotropy). Open and filled symbols denote cases with 
negative and positive velocity gradient imposed, respectively.
{\it Right panel}: the observed relation. The data are taken from those
compiled and analyzed in \citet{Hashimoto13}.
}
\end{figure*}

The left panel of Figure~\ref{fig:EW_Vpeak} summarizes the relation between 
\lya EW (or the apparent luminosity $L/L_0$) and the shift in the (red) 
peak of the 
\lya line. The points are clearly grouped according to the column density
(triangles, squares, and circles for $nR=10^{18}$, $10^{19}$, and 
$10^{20}{\rm cm}^{-2}$, respectively), with larger peak shift at
higher column density. For the sequence at a fixed column density,
the spread in the EW distribution results from the dependence on
the viewing angle. The size of the symbols indicates the magnitude of
the anisotropy of the system, with larger symbols for higher anisotropy.

For the ``density gradient'' case (black points), the anti-correlation 
between EW and $v_{\rm peak}$ is stronger at higher column density, but
overall the anti-correlation is weak. The offset in the peak position is
largely determined by the column density. At fixed column density, a cloud with
a low anisotropy in the initial optical depth (indicated by smaller symbols)
has a smaller spread in the apparent luminosity/EW distribution viewed from
different directions, which is expected. 

For the ``velocity gradient'' case (magenta points), a similar trend of
increasing strength of EW-$v_{\rm peak}$ anti-correlation with increasing
column density is found, but the anti-correlation is much stronger than
the ``density gradient'' case. At fixed velocity gradient, a higher column
density means a lower degree of system anisotropy, and therefore we see a
smaller spread. At fixed column density, a smaller velocity gradient (indicated
by smaller symbols) also means a lower degree of system anisotropy, leading 
to smaller spread in EW.

The extended ``velocity gradient'' case (red points) closely follows the above
results, with filled (open) symbols for imposing positive (negative) velocity
gradients. Similar results are also found for the ``bipolar wind'' case
(blue symbols).

As a whole, for the runs we have performed, the \lya line peak offset is mainly
driven by the column density, with larger offset from systems of higher column 
density. 
The overall trend seen in the left panel 
of Figure~\ref{fig:EW_Vpeak} from all the cases we consider is that the spread
in EW distribution decreases with increasing \lya line peak offset.

Interestingly the above trend appears to be consistent with recent 
observations.  In the right panel of Figure~\ref{fig:EW_Vpeak}, we reproduce 
the data points compiled and analyzed by \citet{Hashimoto13} for LAEs,
\lya blobs (LABs), and Lyman break galaxies (LBGs). The EW can be measured 
from the
\lya line luminosity and continuum with narrow-band and broad-band photometry.
To determine the peak offset, the systemic redshifts of the galaxies are
measured from nebular emission lines such H$\alpha$. 
Even though 
\citet{Hashimoto13} cast their result as an anti-correlation between EW and
$v_{\rm peak}$, the plot suggests a smaller EW spread towards larger peak
offset. We emphasize that we do not intend to fit the observation here, since
our models are simplistic. Nevertheless, it
is encouraging to see that both the ranges of EW and $v_{\rm peak}$ and their
relation fall into the ballpark of the observational results. It suggests that
the key element in our simple models (i.e., \lya emission from anisotropic
systems) could play an important role in shaping the \lya emission in real
systems. More realistic models (e.g., those based on galaxy formation 
simulations) are necessary for a better understanding and comparison with 
the observation.

\begin{figure}
\plotone{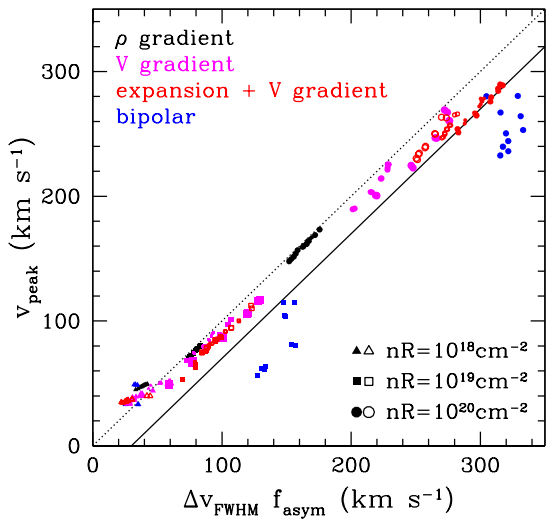}
\caption[]{
\label{fig:VV}
 The relation between the \lya line peak offset $v_{\rm peak}$ and
 the FWHM of the line $\Delta v_{\rm FWHM}$ modified by the line
 asymmetry parameter $f_{\rm asym}$. The asymmetry parameter 
 $f_{\rm asym}$ is defined as three times the ratio of the width at 
 half maximum on the blue and red sides of the line peak. 
 The dotted line denotes the equality of
 the two quantities, while the solid line has a 30$\kms$ downward offset
 to approximate the mean relation. The symbols have the same meanings as 
 in Figure~\ref{fig:EW_Vpeak}.
}
\end{figure}

To study the EW--$v_{\rm peak}$ relation observationally, the quantity
$v_{\rm peak}$ is the more difficult one to measure. The reason is that
a line indicating the systemic redshift of the galaxy is needed. Examples
of such lines are H$\alpha$ and [\ion{O}{3}] 
\citep[e.g.,][]{Steidel10,McLinden11,Hashimoto13}, which shift to infrared 
for high redshift galaxies. It would be
interesting to see whether there are other possible ways that can inform us 
about
$v_{\rm peak}$, without the knowledge of the systemic redshift. Based on our
results, we see a general correlation between $v_{\rm peak}$ and the FWHM
$\Delta v_{\rm FWHM}$, mainly driven by the column density. The degree of
correlation varies from case to case, which leads to a large scatter among
cases with different anisotropy setups. We investigate whether other spectral 
features of the \lya line can help to reduce the scatter. Besides line peak 
offset and the FWHM of the line, another obvious property of the \lya line
is its asymmetry. Both from observation and from our model, the red peak 
of the \lya line is usually asymmetric, having a red tail. We therefore can
introduce an asymmetry parameter. It can be defined by comparing the line fluxes
or widths blueward and redward of the line peak (see, e.g., 
\citealt{Rhoads03}). Here
we define it as three times the ratio of the widths at half maximum at the 
blue and red sides of the peak, $f_{\rm asym}=3W_{\rm blue}/W_{\rm red}$. 
By introducing
this parameter, we find a relatively tight correlation between $v_{\rm peak}$
and $\Delta v_{\rm FWHM} f_{\rm asym}$ (see Figure~\ref{fig:VV}), probably with
the low $v_{\rm peak}$ part of the ``bipolar wind'' model causing the largest
scatter. Note that both $\Delta v_{\rm FWHM}$ and $f_{\rm asym}$ can be 
inferred from the \lya spectra. Observationally, \citet{McLinden13} show
a possible trend of increasing line asymmetry with increasing peak offset. 
It would be interesting to see whether the correlation seen in our study 
holds for realistic models and whether observations support it. Note that
for a fair comparison with observations, the spectral resolution needs to
be taken into account. Its effect is to smooth the \lya line and increase both
the FWHM and the asymmetry parameter, which would cause a slope change 
for the correlation in Figure~\ref{fig:VV}.
In any case, 
a correlation of similar type to ours or with 
similar spirit is worth pursuing. It can not only provide a way to determine 
the peak offset without knowing the systemic redshift, but also serve as a
relation to test theoretical models of environments of star-forming 
galaxies. 

\section{Summary and Discussion}
\label{sec:summary}

We perform a theoretical investigation of the effect of anisotropy in neutral 
hydrogen systems on the observed properties of \lya emission. The motivation
of the work is to help understand the relation between the gas environment 
around star-forming galaxies and \lya emission properties and yield insights
on using the latter to probe the former. We find that the anisotropy in the
spatial and kinematic distributions of neutral hydrogen can play an important 
role in shaping the observed \lya emission properties from star-forming
galaxies. 

We consider simple configurations of neutral hydrogen systems of spherical 
clouds with a central point source of \lya emission. We set up system 
anisotropy induced by different factors. The term ``system anisotropy'' 
refers to the anisotropy in the initial optical depth of \lya photons. 
Since the scattering optical depth of \lya photons depends on both density 
and velocity, we basically explore anisotropies induced by two types of 
causes. The first type is density-caused and a ``density gradient'' is applied 
to the cloud along one direction. The second type is velocity-caused. For the 
``velocity gradient'' case, we apply a velocity gradient along one direction 
to the otherwise static cloud. Its extension, the ``expansion plus velocity 
gradient'' case, has an additional, isotropic Hubble-like expansion. The 
``bipolar wind'' case, which has a spatially separated outflowing region, 
also belongs to this velocity-caused category. For each case, we set up 
systems of different 
column densities and of different degrees of anisotropy. We perform Monte 
Carlo \lya radiative transfer calculations to obtain the \lya emission 
escaping from
the clouds and study the anisotropy in \lya flux and spectral features.

Owing to the resonant scatterings with neutral hydrogen atoms, a \lya photon
takes a random walk in a cloud, with its traveling direction and frequency
constantly changing. Such a random walk enables the photon to explore the
optical depths along different directions. It tends to escape along the
directions of low resistance. An initial anisotropy in \lya scattering 
optical depth therefore translates to the anisotropy in the escaping \lya
emission. It is not surprising that a dipole
(quadrupole) component in \lya flux (or apparent \lya luminosity) is found
for a dipole (quadrupole) component in the system optical depth anisotropy. 
Roughly speaking, the \lya flux in a direction is determined by the optical 
depth along that direction {\it relative} to those along other directions,
or by the fractional excess of the initial optical depth.

For an ensemble of the same systems with random orientations, the 
anisotropic \lya emission gives rise to a non-trivial distribution in the 
observed/apparent \lya luminosity or the \lya EW. The general EW 
distributions from our models are found to be skewed with a tail towards high 
EW values. Such distributions, especially the ones from the ``density 
gradient'' and ``velocity gradient'' cases, resemble the observed ones for 
LAEs \citep[e.g.,][]{Ouchi08,Nilsson09,Ciardullo12}.
The observed \lya EW distribution is likely a superposition of systems of a 
variety of environments. Given the generic features seen in our models, a 
superposition of EW distributions from models with different types of 
anisotropy, different degrees of anisotropy, and different column densities 
will still keep the skewed shape and resemble the observed distributions. 
Even though our models are idealized and hardly realistic, our
results suggest that the viewing angle dependent \lya emission caused by 
system anisotropy can be an important factor in contributing to the EW 
distribution of LAEs.

\lya spectra are also subjected to the anisotropic effect. Noticeably for
a given system, the offset of the line peak depends on the viewing angle. 
\lya photons escaping from the directions of low optical depths typically
have a spectrum with a smaller peak offset from the line center. The overall
scale of the offset depends strongly on the mean column density of the cloud.
The viewing angle dependence of both the apparent luminosity (EW) and peak
offset cause an anti-correlation between the two quantities for a given 
cloud --- along the direction that is easy to escape, we have a higher \lya
flux (apparent luminosity) and \lya photons diffuse less in frequency space.
We note that the peak offset is not necessarily a feature specific 
to the anisotropic systems we study, and it can also occur in isotropic 
systems. In such systems,
the peak offset increases with increasing system column density or optical 
depth. However, for isotropic 
clouds with a given intrinsic \lya luminosity $L_0$, the apparent 
luminosity $L=L_0$, which is independent of the column density. 
Therefore, the anti-correlation between peak offset and luminosity (EW)
we find in the anisotropic systems cannot be naturally produced by the
isotropic systems, unless with the effect of dust invoked or with a 
contrived scenario of putting more luminous sources in systems of lower 
column densities.

For a fixed magnitude of the anisotropy factor (e.g., density or velocity 
gradient), the system anisotropy becomes weaker at higher column density.
From the viewing angle dependence of \lya luminosity or EW at fixed 
column density and the dependence of the peak offset on column density,
we find that at larger peak offset the spread in the apparent luminosity or 
EW is reduced. Interestingly this generic feature between EW and 
peak offset in 
the model anisotropic \lya emission is seen in the observational data in 
\citet{Hashimoto13}, suggesting that anisotropic \lya emission can be 
at work in real \lya emitting systems. 

Based on our simple models, we also find a correlation between the \lya line
peak offset and line shape (i.e., some combination of the FWHM and asymmetry 
of the line). An 
observational test of this correlation could help us understand \lya
radiative transfer in real systems by comparing model predictions and 
observations. Conversely, a similar kind of relation
between \lya peak offset and spectral shape established from 
observation, if possible, would provide opportunities for us to 
learn about the density/velocity structures or, more generally, the environment 
around \lya emitting systems that shapes the \lya emission properties.
Additionally, this type of correlation has the potential use of determining
the systemic redshift of galaxies with the \lya emission line alone.

Our investigation based on analytic models suggests that anisotropies in
the spatial and kinematic distributions of neutral hydrogen in the CGM and 
IGM can be an important ingredient in determining the properties of observed 
\lya emission. While we try to build models that capture some features in 
the CGM and IGM around star-forming galaxies (e.g., density inhomogeneity and 
outflow), they are simplistic and by no means realistic. A further
study along this path is to apply similar analyses to galaxies in high
resolution cosmological galaxy formation simulations. If the relations 
found in this paper turn out to exist for simulated galaxies, 
we expect the scatter to 
be large given the more complex environment around galaxies. 

In fact, some of the features found in our study are broadly similar to 
those from
studies of individual simulated galaxies. For example, the viewing angle
dependent \lya flux \citep[e.g.,][]{Laursen09,Zheng10,Barnes11,Yajima12,
Verhamme12}, correlation of flux with the initial \lya optical depth
(e.g., Figure 11 in \citealt{Zheng10}), and the relation between peak shift
and \lya flux (e.g., Figure 9 in \citealt{Zheng10}). 
Compared to the results from our simple models, the viewing angle dependent 
\lya flux in the above study does show much larger spread, an expected 
consequence of the more complex density and velocity distribution of the gas
around galaxies.

One can certainly extend our simple models to more complicated models, with 
more realistic geometry and coupling between density and velocity, which is 
beyond the scope of this paper. Indeed, models with various setups start to
be studied \citep[e.g.,][]{Laursen13,Behrens14,Gronke14}.
Given what we have learned from the simple models, a complementary route is 
to continue the investigation with 
high-resolution simulated galaxies. A large ensemble of high-resolution 
simulated galaxies are necessary for a statistical study, and 
we reserve such an investigation for future work. From the results and
analyses we present in this paper, we expect that a more detailed study
with simulated galaxies will greatly advance our understanding of the 
interactions between \lya emission and CGM/IGM and of galaxy formation
through the CGM/IGM environment probed by \lya emission.

\acknowledgments

We thank Renyue Cen and Jordi Miralda-Escud{\'e} for useful comments.
This work was supported by NSF grant AST-1208891. J.W. was also
supported by the Undergraduate Research Opportunities Program (UROP) at the
University of Utah.
The support and resources from the Center for High Performance Computing at the University of Utah are gratefully acknowledged.

{}


\begin{thebibliography}{}

\bibitem[Adams(1972)]{Adams72}
Adams, T.~F.\ 1972, \apj, 174, 439

\bibitem[Ahn et al.(2000)]{Ahn00}
Ahn, S.-H., Lee, H.-W., \& Lee, H.~M.\ 2000, Journal of Korean Astronomical Society, 33, 29

\bibitem[Ahn et al.(2001)]{Ahn01}
Ahn, S.-H., Lee, H.-W., \& Lee, H.M. 2001, ApJ, 554, 604

\bibitem[Ahn et al.(2002)]{Ahn02}
Ahn, S.-H., Lee, H.-W., \& Lee, H.M. 2002, ApJ, 567, 922

\bibitem[Ahn \& Lee(2002)]{Ahn02b} 
Ahn, S.-H., \& Lee, H.-W.\ 2002, Journal of Korean Astronomical Society, 35, 
175 

\bibitem[Auer(1968)]{Auer68}
Auer, L.~H.\ 1968, \apj, 153, 783

\bibitem[Avery \& House(1968)]{Avery68}
Avery, L.~W., \& House, L.~L.\ 1968, \apj, 152, 493

\bibitem[Barnes et al.(2011)]{Barnes11} 
Barnes, L.~A., Haehnelt, M.~G., Tescari, E., \& Viel, M.\ 2011, \mnras, 416, 
1723 

\bibitem[Behrens et al.(2014)]{Behrens14}
Behrens, C., Dijkstra, M., \& Niemeyer, J.~C.\ 2014, \aap, 563, A77 

\bibitem[Behrens \& Niemeyer(2013)]{Behrens13}
Behrens, C., \& Niemeyer, J.\ 2013, \aap, 556, A5 

\bibitem[Bland \& Tully(1988)]{Bland88}
Bland, J., \& Tully, B.\ 1988, \nat, 334, 43

\bibitem[Ciardullo et al.(2012)]{Ciardullo12} 
Ciardullo, R., Gronwall, C., Wolf, C., et al.\ 2012, \apj, 744, 110 

\bibitem[Dekel et al.(2009)]{Dekel09} 
Dekel, A., Birnboim, Y., Engel, G., et al.\ 2009, \nat, 457, 451 

\bibitem[Dijkstra et al.(2006)]{Dijkstra06} 
Dijkstra, M., Haiman, Z., \& Spaans, M.\ 2006, \apj, 649, 14 

\bibitem[Dijkstra, Lidz, \& Wyithe(2007)]{Dijkstra07}
Dijkstra, M., Lidz, A., \& Wyithe, J.~S.~B.\ 2007, \mnras, 377, 1175

\bibitem[Gawiser et al.(2007)]{Gawiser07} 
Gawiser, E., et al.\ 2007, \apj, 671, 278 

\bibitem[Gronke \& Dijkstra(2014)]{Gronke14} 
Gronke, M., \& Dijkstra, M.\ 2014, arXiv:1406.6709 

\bibitem[Guaita et al.(2010)]{Guaita10}
Guaita, L., Gawiser, E., Padilla, N., et al.\ 2010, \apj, 714, 255 

\bibitem[Harrington(1973)]{Harrington73}
Harrington, J.~P.\ 1973, \mnras, 162, 43

\bibitem[Hashimoto et al.(2013)]{Hashimoto13} 
Hashimoto, T., Ouchi, M., Shimasaku, K., et al.\ 2013, \apj, 765, 70 

\bibitem[Hill et al.(2008)]{Hill08} 
Hill, G.~J., et al.\ 2008, Astronomical Society of the Pacific Conference
Series, 399, 115 

\bibitem[Kere{\v s} et al.(2005)]{Keres05} 
Kere{\v s}, D., Katz, N., Weinberg, D.~H., \& Dav{\'e}, R.\ 2005, \mnras, 363, 
2 

\bibitem[Laursen et al.(2009)]{Laursen09}
Laursen, P., Sommer-Larsen, J., \& Andersen, A.~C.\ 2009, \apj, 704, 1640 

\bibitem[Laursen et al.(2011)]{Laursen11} 
Laursen, P., Sommer-Larsen, J., \& Razoumov, A.~O.\ 2011, \apj, 728, 52 

\bibitem[Laursen et al.(2013)]{Laursen13}
Laursen, P., Duval, F., {\"O}stlin, G.\ 2013, \apj, 766, 124 

\bibitem[Loeb \& Rybicki(1999)]{Loeb99}
Loeb, A., \& Rybicki, G.~B.\ 1999, \apj, 524, 527 

\bibitem[Malhotra \& Rhoads(2002)]{Malhotra02} 
Malhotra, S., \& Rhoads, J.~E.\ 2002, \apjl, 565, L71 

\bibitem[McLinden et al.(2011)]{McLinden11} 
McLinden, E.~M., Finkelstein, S.~L., Rhoads, J.~E., et al.\ 2011, \apj, 730, 
136 

\bibitem[McLinden et al.(2013)]{McLinden13} 
McLinden, E.~M., Malhotra, S., Rhoads, J.~E., et al.\ 2013, \apj, 767, 48 

\bibitem[Neufeld(1990)]{Neufeld90}
Neufeld, D.~A.\ 1990, \apj, 350, 216

\bibitem[Nilsson et al.(2009)]{Nilsson09}
Nilsson, K.~K., Moeller-Nilsson, O., Moeller, P., Fynbo, J.~P.~U.,
\& Shapley, A.~E.\ 2009, \mnras, 400, 232

\bibitem[Noterdaeme et al.(2012)]{Noterdaeme12} 
Noterdaeme, P., Laursen, P., Petitjean, P., et al.\ 2012, \aap, 540, A63 

\bibitem[Ouchi et al.(2008)]{Ouchi08} 
Ouchi, M., et al.\ 2008, \apjs, 176, 301

\bibitem[Partridge \& Peebles(1967)]{Partridge67}
Partridge, R.B., \& Peebles, P.J.E. 1967, ApJ, 147, 868

\bibitem[Rhoads et al.(2003)]{Rhoads03}
Rhoads, J.E., Dey, A., Malhotra, S., Stern, D., Spinrad, H., Jannuzi, B.T., Dawson, S., Brown, M.J.I., \& Landes, E. 2003, AJ, 125, 1006

\bibitem[Roy et al.(2009)]{Roy09} 
Roy, I., Qiu, J.-M., Shu, C.-W., \& Fang, L.-Z.\ 2009, New Astronomy, 14, 513 

\bibitem[Roy et al.(2010)]{Roy10}
Roy, I., Shu, C.-W., \& Fang, L.-Z.\ 2010, \apj, 716, 604 

\bibitem[Rubin et al.(2013)]{Rubin13} 
Rubin, K.~H.~R., Prochaska, J.~X., Koo, D.~C., et al.\ 2013, arXiv:1307.1476 

\bibitem[Shapley et al.(2003)]{Shapley03} 
Shapley, A.~E., Steidel, C.~C., Pettini, M., \& Adelberger, K.~L.\ 2003, \apj, 
588, 65 

\bibitem[Shopbell \& Bland-Hawthorn(1998)]{Shopbell98}
Shopbell, P.~L., \& Bland-Hawthorn, J.\ 1998, \apj, 493, 129

\bibitem[Steidel et al.(2010)]{Steidel10} 
Steidel, C.~C., Erb, D.~K., Shapley, A.~E., et al.\ 2010, \apj, 717, 289 

\bibitem[Veilleux \& Rupke(2002)]{Veilleux02}
Veilleux, S., \& Rupke, D.~S.\ 2002, \apjl, 565, L63

\bibitem[Verhamme et al.(2006)]{Verhamme06}
Verhamme, A., Schaerer, D., \& Maselli, A.\ 2006, \aap, 460, 397 

\bibitem[Verhamme et al.(2012)]{Verhamme12} 
Verhamme, A., Dubois, Y., Blaizot, J., et al.\ 2012, \aap, 546, A111 

\bibitem[Weiner et al.(2009)]{Weiner09} 
Weiner, B.~J., Coil, A.~L., Prochaska, J.~X., et al.\ 2009, \apj, 692, 187 

\bibitem[Wyithe \& Dijkstra(2011)]{Wyithe11} 
Wyithe, J.~S.~B., \& Dijkstra, M.\ 2011, \mnras, 415, 3929 

\bibitem[Yajima et al.(2012)]{Yajima12} 
Yajima, H., Li, Y., Zhu, Q., et al.\ 2012, \apj, 754, 118 

\bibitem[Zheng \& Miralda-Escud{\'e}(2002)]{Zheng02} 
Zheng, Z., \& Miralda-Escud{\'e}, J.\ 2002, \apj, 578, 33 

\bibitem[Zheng et al.(2010)]{Zheng10} 
Zheng, Z., Cen, R., Trac, 
H., \& Miralda-Escud{\'e}, J.\ 2010, \apj, 716, 574 

\bibitem[Zheng et al.(2011a)]{Zheng11a} 
Zheng, Z., Cen, R., Trac, 
H., \& Miralda-Escud{\'e}, J.\ 2011, \apj, 726, 38 

\bibitem[Zheng et al.(2011b)]{Zheng11b} 
Zheng, Z., Cen, R., 
Weinberg, D., Trac, H., \& Miralda-Escud{\'e}, J.\ 2011, \apj, 739, 62 



\end{thebibliography}
\end{document}